\documentclass[
reprint,
 amsmath,amssymb,
aps,
prb,
]{revtex4-2}

\usepackage{graphicx}
\usepackage{dcolumn}
\usepackage{bm}
\usepackage{hyperref}
\usepackage[mathlines]{lineno}

\usepackage{svg}
\usepackage{braket}
\usepackage[version=4]{mhchem}
\usepackage[super]{nth}

\usepackage{isomath}
\usepackage{tikz}

\newcommand{\myvect}[1]{\vectorsym{#1}}
\newcommand{\vect}[1]{\myvect{#1}}

\newcommand{\of}[1]{\left(#1\right)}
\newcommand{\off}[1]{ \left[ #1 \right] }
\newcommand{\offf}[1]{\left\{#1\right\}}
\newcommand{\abss}[1]{\left|#1\right|}

\newcommand{\expec}[1]{\left<#1\right>}

\newcommand{\tm}[1]{\text{#1}}
\newcommand{\ofl}[1]{\left(#1\right.}
\newcommand{\ofr}[1]{\left.#1\right)}

\newcommand{\matt}[1]{\text{#1}}

\newcommand{\diff}{\mathop{}\!\mathrm{d}}

\newcommand{\emit}{\mbox{Ir(ppy)\textsubscript{2}acac}}
\newcommand{\hostc}{\mbox{m-MTDATA\textsuperscript{+}}}
\newcommand{\host}{\mbox{m-MTDATA}}

\usepackage{xargs}                      

\begin{document}

\preprint{arXiv}

\title{Dipole-quadrupole coupling in triplet exciton-polaron quenching in a phosphorescent OLED emission layer} 
\thanks{This paper has been published in Physical Review B \cite{vanhoeselDipolequadrupoleCouplingTriplet2025}.}%

\author{Clint van Hoesel}
\author{Reinder Coehoorn}%
\author{Peter Bobbert}%
 \email{p.a.bobbert@tue.nl}
\affiliation{%
 Department of Applied Physics and Science Education\\Eindhoven University of Technology\\P.O. Box 513, 5600 MB, Eindhoven, The Netherlands
}%
\affiliation{%
 Institute for Complex Molecular Systems\\Eindhoven University of Technology\\P.O. Box 513, 5600 MB, Eindhoven, The Netherlands 
}%

\collaboration{SEQUOIA project} \noaffiliation

\date{\today}

\keywords{organic light-emitting diode, phosphorescent emitter, triplet-polaron quenching, Förster transfer, dipole-quadrupole coupling}

\begin{abstract}
    Improving the efficiency and stability of organic light-emitting diodes (OLEDs) will further expand their present success in display applications. Triplet exciton-polaron quenching (TPQ) is an important cause of limited efficiency and stability in modern phosphorescent OLEDs, where triplet excitons are the emitting species. Lack of understanding of the TPQ mechanism in these OLEDs impedes the development of more efficient and stable OLEDs. We investigate the TPQ mechanism for triplet excitons on a phosphorescent guest interacting with hole polarons on a host. Our quantum-chemical calculations show that at distances relevant for TPQ the Förster approximation for the TPQ rate fails and that dipole-quadrupole coupling is dominant. This resolves a discrepancy between estimates of the TPQ rate obtained from an OLED device study and from the overlap between the emission spectrum of the emitter and absorption spectrum of the charged host. Equivalently to the Förster radius for dipole-dipole TPQ, the dipole-quadrupole TPQ rate can be quantified by a dipole-quadrupole radius obtained from the overlap between the emission spectrum of the emitter and the quadrupolar absorption spectrum of the charged host. The findings of this work are expected to have a broad relevance and to be useful in developing phosphorescent emitter-host combinations with reduced TPQ.
\end{abstract}

\maketitle

\section{\label{sec:Introduction}Introduction}

Developing a full understanding of excitonic processes is key for optimizing the functioning of organic light-emitting diodes (OLEDs), which currently constitute the largest portion of the commercial display market \cite{bauriRecentAdvancesEfficient2021,hongBriefHistoryOLEDs2021}. The presently most efficient OLED emission layers consist of a phosphorescent emitter as guest, embedded in a host. The phosphorescent emitter contains heavy metal atoms for inducing spin-orbit coupling \cite{baldoHighlyEfficientPhosphorescent1998}. The spin-orbit coupling allows triplets to decay radiatively, thus making it in principle possible to achieve 100\% conversion of electron-hole pairs to photons \cite{kawamura100PhosphorescenceQuantum2005}. In contrast, emission based on fluorescence allows only singlet excitons to decay radiatively, leading to a much lower efficiency. However, because the lifetime of triplet excitons on a phosphorescent emitter is long as compared to that of singlet excitons on a fluorescent emitter ($\sim$$\mu$s vs $\sim$ns), there is a comparatively large probability that triplet excitons undergo a quenching process instead of radiative decay, leading to an efficiency loss. 

Both triplet-triplet annihilation (TTA) and triplet-polaron quenching (TPQ) have been identified as important efficiency loss processes \cite{murawskiEfficiencyRollOffOrganic2013}. In addition, these processes can lead to chemical degradation and low operational lifetimes \cite{giebinkDirectEvidenceDegradation2009,choRootCausesLimited2016,laaperiOLEDLifetimeIssues2008,zhangDegradationMechanismsBlue2017}. Regarding TPQ, it has been noted that charged molecules have weakened bonds \cite{wangNegativeChargeManagement2021,kimMobilityBalanceLightemitting2017}, making them susceptible to degradation. It has been suggested that for many systems TPQ is the dominant contribution to the efficiency loss and degradation \cite{vaneerselMonteCarloStudy2014,regnatInfluenceBiasdependentEmission2019}. 
In TPQ, a triplet exciton on a phosphorescent emitter molecule interacts with an (electron or hole) polaron on either an emitter or a host molecule, leading to transfer of the triplet energy to the molecule carrying the polaron and therefore quenching of the triplet exciton.
TPQ has been studied by time-resolved photoluminescence (TRPL) experiments \cite{reinekeTripletexcitonQuenchingOrganic2007,wehrmeisterCombinedElectricalOptical2015,schmidtAnalyzingDegradationEffects2015,ligthartMechanisticDescriptionEfficiency2021,yangInterfacialExcitonpolaronQuenching2024}. In these experiments,  a device under operation (so that polarons are present) is illuminated by a laser pulse, after which the transient photoluminescence (PL) is measured. A TPQ rate is then obtained from a fit to the measured transient PL with a low-dimensional rate equation model  \cite{reinekeTripletexcitonQuenchingOrganic2007,wehrmeisterCombinedElectricalOptical2015,schmidtAnalyzingDegradationEffects2015,yangInterfacialExcitonpolaronQuenching2024} or a microscopic three-dimensional (3D) model \cite{ligthartMechanisticDescriptionEfficiency2021}.

\begin{figure}
    \centering
    \includegraphics[width=8cm]{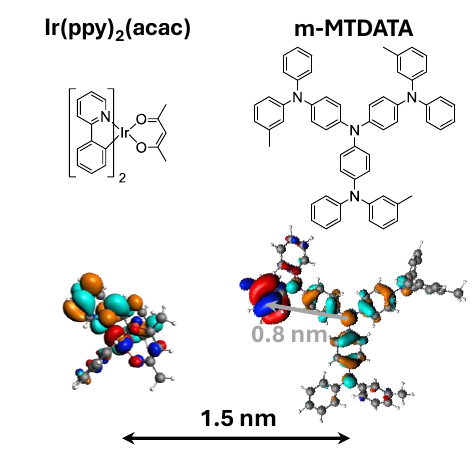}
    \caption{\textbf{
    Excitations on the emitter and the positively charged host involved in TPQ.} Top: chemical structures of the emitter {\emit} and the host {\host}. Bottom: isocontour plots of the occupied (red and blue) and unoccupied (orange and cyan) natural transition orbitals (NTOs) of the excitations of {\emit} and {\hostc}, involved in triplet-polaron quenching (TPQ) between a triplet exciton on {\emit} and a hole polaron on {\host}. The molecules have been drawn on the same scale at a $1.5$ nm center-of-mass distance. The excitation of {\hostc} involves intramolecular transfer of a hole over a distance of about $0.8$ nm (grey arrow).}
    \label{fig:Figuur_2_mMTDATA_Irppy2acac}
\end{figure}

In the present work, we consider TPQ involving hole polarons that are confined to the host. This is the case when the host has a lower ionization potential than the emitter.
In such a case the transport of the holes is only very slightly influenced by the presence of the emitter. As a result, the rate with which polarons encounter excitons can be deduced relatively easily from a \mbox{charge-transport} study for a pure host device. Making use of such a charge-transport study, TPQ has recently been studied experimentally for  
the host-guest blend with {\host} (\mbox{4,4',4''-tris[(3-methylphenyl)phenylamino]triphenylamine}) as host and 7 wt\% {\emit} (\mbox{bis(2-phenylpyridine)(acetylacetonate)iridium(III)}) as emitter guest \cite{ligthartMechanisticDescriptionEfficiency2021}. The chemical structures of {\host} and {\emit} are shown in \textbf{Figure~\ref{fig:Figuur_2_mMTDATA_Irppy2acac}} (top). The hole transport was investigated in hole-only devices of {\host} and modelled with 3D kinetic Monte Carlo (kMC) simulations to determine the parameters for hole transport. The hole transport in a hole-only {\host}:{\emit} device was indeed found to be almost identical to that in the hole-only {\host} devices \cite{ligthartMechanisticDescriptionEfficiency2021}. The measured decay of the PL efficiency and the PL lifetime as a function of current density in a TRPL experiment could be modelled well with 3D kMC simulations, using the parameters for hole transport obtained for the hole-only {\host} devices and assuming TPQ to be a Förster-type process with an effective Förster radius $R_{\tm{F}}=3.8$ nm \cite{ligthartMechanisticDescriptionEfficiency2021}. We will call this method of obtaining the Förster radius the Förster-fit (FF) method.

If TPQ in the {\host}:{\emit} blend would indeed be a Förster-type process, then the Förster radius could also be obtained from the overlap between the emission spectrum of the emitter and the absorption spectrum of the positively charged host \cite{foersterZwischenmolekulareEnergiewanderungUnd1948}. We will call this method of calculating the Förster radius the spectral overlap (SO) method. Calculating the Förster radius in this way from the measured emission spectrum of {\emit} in a thin host-guest film and the absorption spectrum of {\hostc} measured in a spectroelectrochemistry experiment in solution, a value \mbox{$R_{\tm{F}}=3.1$ nm} is obtained \cite{jasparsSpectroElectrochemicalDetermination2021}. This value differs strongly from the value of 3.8 nm obtained from the FF method. 
Since a Förster-type rate scales as $R_{\tm{F}}^{6}$, this difference translates into a large difference of $\of{3.8/3.1}^6 \approx 3.4$ in the TPQ rates.
 
In the present paper, we study the origin of the difference in Förster radii found with the FF and SO methods by performing advanced quantum-chemical calculations of the TPQ process that include vibrational and environmental effects. We investigate the possibility that the difference is caused by the different environments of an {\hostc} molecule in the thin film and in solution. However, we come to the conclusion that this is not the case. By performing exact direct Coulomb integral calculations of the coupling between a triplet exciton on an {\emit} molecule and a hole on an {\host} molecule, we find that the TPQ rate has an important dipole-quadrupole contribution.
As due to diffusion of hole polarons over the {\host} matrix their distance to the triplet excitons can become very small, the \mbox{dipole-quadrupole} contribution to the TPQ rate can become very large. We will argue that a relevant distance for TPQ is about $1.5$ nm. In Figure~\ref{fig:Figuur_2_mMTDATA_Irppy2acac} (bottom) we show an {\emit} molecule and an {\hostc} molecule separated by $1.5$ nm, together with the natural transition orbitals (NTOs) that have the strongest contribution to the excitations involved in the TPQ. The NTOs on the {\emit} molecule show the metal-to-ligand electron transfer typical for Ir-based emitters. The NTOs on the {\hostc} molecule show an intramolecular transfer of the hole from the triphenylamine core to one of the methylphenyl side groups over a distance of about \mbox{$0.8$ nm}. It is clear from the figure that the distance over which the hole is transferred is not small in comparison to the distance between the molecules. This indicates that the Förster dipole-dipole approximation may fail and that higher-order multipole contributions to the TPQ rate should be considered. We will show in this paper that including the dipole-quadrupole contribution can, at least for a large part, account for the discrepancy between the Förster radii found with the FF and the SO method in this system.

\section{\label{sec:Results}Results}
The emission spectrum of iridium-cored emitters like {\emit} can be easily measured and is not very sensitive to the specific environment of this emitter.
Therefore, we can, in the TPQ process involving a triplet exciton on {\emit} and a hole polaron on {\host}, focus on the absorption spectrum of positively charged {\host} molecules ({\hostc}). In contrast to the {\emit}, the {\host} molecule is rather extended and flexible, such that its properties could be expected to depend on its conformation and interaction with the environment.

\begin{figure*}
    \centering
    \includegraphics[width=11cm]{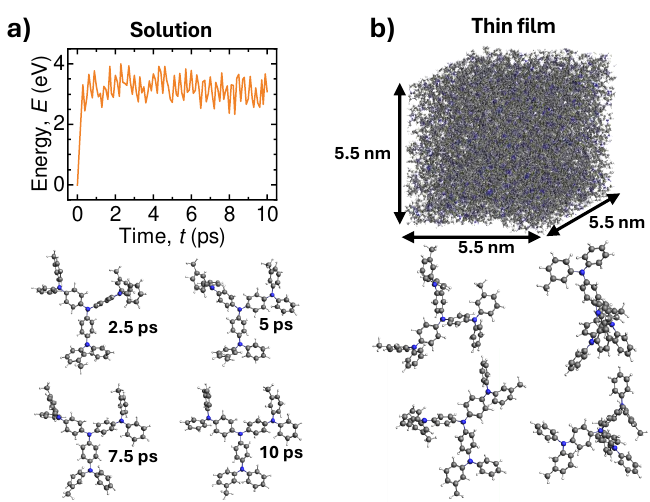}
    \caption{\textbf{Simulated conformers of host molecules in solution and the thin film.} (a) Top: energy of an {\hostc} molecule during a molecular dynamics simulation in solution with the COSMO embedding model with $\varepsilon_{\tm{r}} = 3$. Bottom: four conformers at the indicated times in the simulation. (b) Top: $5.5$ nm thick middle slice of a $5.5 \times 5.5 \times 14.7$ nm box of {\host} molecules in a thin film, generated by the AMS OLED deposition workflow \cite{rugerSCMOLEDDeposition}. Bottom: $4$  randomly chosen conformers from the slice.}
    \label{fig:Figuur_1_workflowv2}
\end{figure*}

To study the absorption spectrum of solvated 
positively charged {\host} molecules, 
we first perform a molecular dynamics (MD) simulation of an {\hostc} molecule in solution, taking on different conformations (conformers) as a function of time. Results of the simulation are shown in \textbf{Figure~\ref{fig:Figuur_1_workflowv2}}(a). The effects of the environment were taken into account via the COSMO solvation model as implemented in the AMS software \cite{teveldeChemistryADF2001,pyeImplementationConductorlikeScreening1999}. The relative dielectric constant of the environment was set to $\varepsilon_{\tm{r}}=3$, a typical approximate value for organic semiconductors and several organic solvents. This value will be used for all calculations with COSMO solvation. Further details are given in Section \ref{subsec:solutionfilmworkflow} in \hyperref[sec:Methods]{Methods}. 
The top part of Figure~\ref{fig:Figuur_1_workflowv2}(a) displays the energy of the {\hostc} molecule as a function of time, starting from the \mbox{lowest-energy} molecular conformation in the gas phase, which provides our reference energy $E=0$. We observe that after about $0.5$ ps the system has equilibrated. The lower part of Figure~\ref{fig:Figuur_1_workflowv2}(a) shows examples of conformers that are sampled at the indicated times.

\begin{figure}
    \centering
    \includegraphics[width=8.5cm]{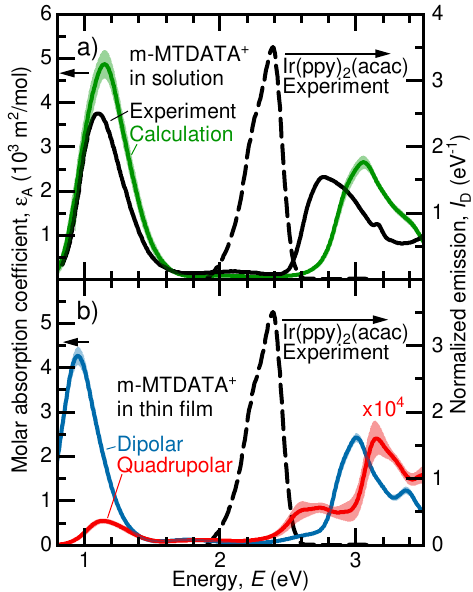}
    \caption{\textbf{Absorption spectra of the positively charged host and emission spectrum of the emitter.} a) Full black line: absorption spectrum of positively charged {\hostc} measured in a DCM solution  \cite{jasparsSpectroElectrochemicalDetermination2021}. Green line: theoretical absorption spectrum in solution ($\varepsilon_{\tm{r}}=3$), calculated from an average of $95$ conformers taken from the solution in Figure~\ref{fig:Figuur_1_workflowv2}(a). b) Blue line: dipolar contribution to the calculated absorption spectrum of {\hostc} in a thin film, averaged over 163 conformers taken from the thin film in Figure~\ref{fig:Figuur_1_workflowv2}(b). Red line: quadrupolar contribution to the absorption spectrum, averaged over 21 conformers taken from the thin film. Shaded bands in a) and b): $2 \sigma$ confidence interval. Dashed black line in a) and b): normalized emission spectrum of 7 wt\% {\emit} in a thin film blend with {\host} \cite{ligthartEffectTripletConfinement2018}.}
    \label{fig:ExperimentalSpectraOverlap}
\end{figure}

The green line in \textbf{Figure~\ref{fig:ExperimentalSpectraOverlap}}(a) shows the theoretical absorption spectrum of {\hostc} in solution, calculated with time-dependent density functional theory (TD-DFT), accounting for vibrational effects. The spectrum was calculated by taking the average of the individual absorption spectra of $95$ {\hostc} molecular conformers from the simulation of Figure~\ref{fig:Figuur_1_workflowv2}(a), sampled with equal time intervals of $0.1$ ps, after the equilibration time of $0.5$ ps. The shading indicates the $2\sigma$ confidence interval of the mean of the spectra. Details of the calculation of the spectra can be found in Section \ref{subsec:solutionfilmworkflow} in \hyperref[sec:Methods]{Methods}. 
Supplemental Figure~\ref{fig:IndividualDipoleAbsorptionSpectraSolution} \cite{SupplementaryInformation} shows the absorption spectra of the $95$ individual conformers. 
The full black line in the figure shows the molar absorption spectrum of {\hostc} in a \mbox{dichloromethane} (DCM) solution, measured in a spectroelectrochemistry experiment \cite{jasparsSpectroElectrochemicalDetermination2021}. 
We conclude that there is a fairly good agreement between the calculated and measured absorption spectrum of {\hostc} in solution. We attribute the differences to approximations in the \mbox{TD-DFT} calculations and to the differences in the dielectric embedding. The refractive index $n$ of DCM is $1.4242$ at \mbox{$\lambda = 589$ nm} ($2.11$ eV) \cite{CRCHandbook}, leading to a relative dielectric constant $\varepsilon_{\tm{r}} = n^2 = 2.0283$, which differs from the value $\varepsilon_{\tm{r}} = 3$ used in the COSMO embedding. The high-energy peak in the spectra at about $3$ eV corresponds mainly 
to excitations across the gap 
(the onset of the experimental optical absorption spectrum of {\host} is $3.1$ eV \cite{gravesPhotophysicalInvestigationThermally2014}). The low-energy peak at around $1$ eV is mainly caused by excitations from HOMO-1 and HOMO-2 to the singly occupied molecular orbital (SOMO) of {\hostc}.

The dashed line in Figure~\ref{fig:ExperimentalSpectraOverlap}(a) shows the normalized measured emission spectrum of a $50$ nm thin film of 7 wt\% {\emit} in {\host} \cite{ligthartEffectTripletConfinement2018}.
This emission spectrum is almost indistinguishable from the measured emission spectra of {\emit} in other host materials 
\cite{ligthartEffectTripletConfinement2018}, demonstrating that environmental effects are insignificant for the emission of {\emit}.
The 7 wt\% emitter concentration is for Ir-cored emitters still in the range where there is no significant effect of the concentration on the emission spectrum \cite{kawamuraIntermolecularInteractionConcentrationQuenching2006}, so that we can treat the emission spectrum as that of a collection of independent emitters. This also suggests that self-absorption, i.e., the absorption of emitted photons by the emission layer itself, is negligible.
In Supplemental Figure~\ref{fig:AutoAbsorption} the effect of the explicitly calculated self-absorption on the emission spectrum is shown and found to be negligible.
The emission peak of {\emit} lies in between the low-energy and high-energy peaks in the absorption spectrum of {\hostc}. Nevertheless, there is an overlap with the {\hostc} absorption spectrum that can give rise to TPQ. The absorption spectrum in the overlap region ($\sim$ 2-2.5 eV) is mainly due to excitations from energetically deeper lying HOMOs than HOMO-1 and HOMO-2 to the SOMO of {\hostc}. The natural transition orbitals (NTOs) of one such excitation, involving the transfer of the hole from the triphenylamine core to one of the methylphenyl side groups of the molecule, are shown in Figure~\ref{fig:Figuur_2_mMTDATA_Irppy2acac} (bottom right). These excitations have lower transition dipole moments than the energetically lower-lying (around $1$ eV) and higher-lying (around $3$ eV) excitations because of the small overlap between the hole and electron NTOs, as observed in Figure~\ref{fig:Figuur_2_mMTDATA_Irppy2acac} (bottom right), leading to a smaller absorption coefficient. 

The rate for Förster transfer of an exciton on a donor molecule D to an acceptor molecule A is \cite{kohlerElectronicProcessesOrganic2015a}
\begin{align}
    k_{\matt{F}} \of{r} =& \frac{9 \ln \of{10} \eta\expec{\kappa^2} }{128 \pi^5 n^4 N_{\matt{A} } \tau  r^6} \int_{0}^{\infty} I_{\matt{D}} \of{\lambda} \varepsilon_{\matt{A}} \of{\lambda} \lambda^4 \, \diff{\lambda} \nonumber\\
     =& \frac{9 c^4 \hbar^2 \ln \of{10} \eta\expec{\kappa^2} }{8 \pi n^4 N_{\matt{A} } \tau r^6} \int_{0}^{\infty} \frac{I_{\matt{D}} \of{E} \varepsilon_{\matt{A}} \of{E} }{ E^{4} } \,  \diff{E} \nonumber\\
    \equiv& \frac{1}{\tau_\matt{r}} \of{\frac{R_{\matt{F}}}{r}}^6\label{eq:BasicFoersterFormula},
\end{align}
with $\eta$ the PL efficiency, $\expec{\kappa^2}$ the average square of the dipole-dipole orientation factor, $n$ the refractive index, $N_{\tm{A}}$ Avogadro's constant, $\tau_{\tm{r}}$ the radiative lifetime of the exciton, $r$ the distance between the molecules, $\lambda$ the wavelength of the light, $I_{\tm{D}} \of{\lambda}$ the normalized emission spectrum of the donor ({\emit}), $\epsilon_{\tm{A}} \of{\lambda}$ the molar absorption coefficient of the acceptor ({\hostc}), \mbox{$\tau=\tau_{\tm{r}}  \eta$} the total lifetime of the exciton, including non-radiative decay, and $R_{\tm{F}}$ the Förster radius.
In the second form of Equation~\ref{eq:BasicFoersterFormula} a transformation is made from wavelength $\lambda$ to energy $E$ in the integration.
We take \mbox{$\expec{\kappa^2}=2/3$}, corresponding to dynamic isotropic averaging \cite{vandermeerKappasquaredNuisanceNew2002}. This averaging is valid because in our case polaron hopping is fast as compared to the exciton decay. This may be understood as follows. From the experimentally determined room-temperature hole mobility of {\host}, $\mu = 3 \times 10^{-9}$ m\textsuperscript{2}/V  \cite{ligthartMechanisticDescriptionEfficiency2021}, we find using the Einstein relation $D=\mu k_{\rm B}T/e$ (where $k_{\rm B}T$ is the thermal energy and $e$ the unit charge) a diffusion coefficient $D\approx 8 \times 10^{-11}$ m\textsuperscript{2}/s. This corresponds to an average hopping rate of about $5 \times 10^{8}$ s\textsuperscript{-1} between nearest-neighbour molecules at a typical distance $r_0\approx 1$ nm, which is much larger than the exciton decay rate \mbox{$\tau^{-1}=6.3 \times 10^{5}$ s\textsuperscript{-1}}, as obtained from the experimental lifetime $\tau=1.6$ $\mu$s \cite{lamanskySynthesisCharacterizationPhosphorescent2001}. 
Hence, many host molecules with different orientations are visited by a polaron during the lifetime of an exciton. 
This also justifies our use of the average of the absorption spectra of many conformers in the theoretical result shown in Figure~\ref{fig:ExperimentalSpectraOverlap}(a).

Applying Equation~\ref{eq:BasicFoersterFormula} to the overlap of the experimental emission spectrum of {\emit} with the experimental and theoretical absorption spectra of {\hostc} in Figure~\ref{fig:ExperimentalSpectraOverlap}(a), we find $R_\tm{F} = 3.1$ nm (the value found in Ref.~\cite{jasparsSpectroElectrochemicalDetermination2021}) and $2.7$ nm, respectively. The agreement between these two values is reasonable. The difference is caused by the difference between the experimental and theoretical absorption spectrum of {\hostc}, which, as we mentioned above, can be attributed to approximations in the calculation of the theoretical absorption spectrum. These approximations lead to an underestimation of the oscillator strength in the spectral overlap of about a factor of a half (compare the green to the black line in the overlap region of the figure). 
However, both values $R_{\tm{F}} = 3.1$ and $2.7$ nm found from the \mbox{spectral-overlap} (SO) method disagree strongly with the Förster radius $R_{\tm{F}} = 3.8$ nm found in the Förster-fit (FF) analysis \cite{ligthartMechanisticDescriptionEfficiency2021}. We note that the TPQ rate is proportional to $R_{\tm{F}}^6$, so that the corresponding rate in the device experiments of Ref.~\cite{ligthartMechanisticDescriptionEfficiency2021} is much larger than found using the measured absorption spectrum in the spectroelectrochemical experiment in Ref.~\cite{jasparsSpectroElectrochemicalDetermination2021} or using our theoretical absorption spectrum (a factor \mbox{$\of{3.8/3.1}^6 \approx 3.4$} for the experimental case to \mbox{$\of{3.8/2.7}^6 \approx 7.8$} for the theoretical case). It is important to establish the reason for this large discrepancy in order to find emitter-host combinations with a suppressed TPQ.

In Ref.~\cite{jasparsSpectroElectrochemicalDetermination2021} the {\hostc} absorption spectrum was measured in solution, whereas the measurements of Ref.~\cite{ligthartMechanisticDescriptionEfficiency2021} were performed for a thin film. 
In order to investigate whether the above discrepancy is caused by this difference, we repeated the calculations of the absorption spectrum of {\hostc} for a thin film, using the AMS OLED deposition and properties workflow \cite{rugerSCMOLEDDeposition}.
The top part of Figure~\ref{fig:Figuur_1_workflowv2}(b) shows a morphology of a thin film of {\host} molecules generated with this workflow. 
The bottom part of Figure~\ref{fig:Figuur_1_workflowv2}(b) shows four randomly chosen molecules from the thin film. Due to steric hindrance, the molecules exhibit stronger conformational deviations from the lowest-energy vacuum conformation than in solution, as can be seen by comparing the corresponding conformers. Because in phosphorescent emitter layers the emitter guest is embedded in the host with only a small concentration, we expect that our calculations of environmental effects of {\host} molecules in a pure {\host} film will be very similar to those of {\host} molecules in the actual blend.

The blue line in Figure~\ref{fig:ExperimentalSpectraOverlap}(b) shows the calculated absorption spectrum of {\hostc} molecules in a thin film, based on calculations for $163$ conformers taken from the middle slice of the simulated film, using the AMS OLED workflow (Supplemental Figure~\ref{fig:IndividualDipoleAbsorptionSpectraMorphology}  \cite{SupplementaryInformation} shows the absorption spectra of the individual conformers). In the workflow, the influence of molecules surrounding a particular conformer is taken into account with the DRF QM/MM embedding method \cite{jensenDiscreteSolventReaction2003}
(see Section \ref{subsec:thinfilmworkflow} in \hyperref[sec:Methods]{Methods} for details). We observe that the calculated absorption spectrum in the film is very similar to the calculated absorption spectrum in solution (green line in Figure~\ref{fig:ExperimentalSpectraOverlap}(a)). The small redshift of the whole spectrum can be attributed to the difference in the modeled embedding between solution and thin film. The Förster radius calculated with Equation~\ref{eq:BasicFoersterFormula} from the overlap between the theoretical thin-film absorption spectrum of {\hostc} and the experimental emission spectrum of {\emit} is within the uncertainty equal to that found in solution: $R_{\tm{F}}=2.7$ nm. We thus conclude that conformational differences and differences in the surrounding of an {\hostc} molecule between the thin film and solution do not resolve the discrepancy between the Förster radii found with the FF and the SO method, so that the discrepancy should be attributed to another effect.

If TPQ would be solely governed by Förster transfer with an $r^{-6}$ distance dependence, then the average distance for a TPQ event in the case of fast polaron hopping can be roughly approximated as
\begin{equation}\label{eq:TypicalForsterInteractionLength}
    \expec{r} \approx \frac{\int_{r_0}^{\infty} r\, r^{-6} r^2\diff{r} }{ \int_{r_0}^{\infty} r^{-6} r^2\diff{r} }=\frac{\int_{r_0}^{\infty} r^{-3} \diff{r} }{ \int_{r_0}^{\infty} r^{-4} \diff{r} } = \frac{3}{2} r_0,
\end{equation}
where again use is made of the fact that polaron hopping is fast and where we have for simplicity assumed that the host can be described as a continuum beyond the nearest-neighbour distance $r_0$. With $r_0 \approx 1$ nm for the typical nearest-neighbour distance, we obtain $\expec{r} \approx 1.5$ nm. From the spatial structure of the corresponding NTOs on the {\emit} and {\hostc} molecules, one can see in Figure~\ref{fig:Figuur_2_mMTDATA_Irppy2acac} that the excitation on the {\hostc} molecule that is relevant for TPQ involves an intramolecular charge transfer over a distance that is not small with respect to $\expec{r}$. This is an indication that the dipole-dipole approximation underlying Förster transfer may fail and that higher-order multipole interactions should be considered.

\begin{figure}
    \centering
    \includegraphics[width=8.5cm]{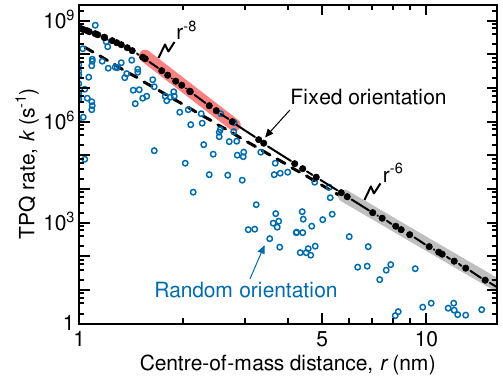}
    \caption{\textbf{TPQ rates as function of emitter-host distance.} Direct Coulomb TPQ rates for the lowest energy gas-phase conformers of {\emit} and {\hostc}. Black closed symbols connected by solid line: single fixed orientation. Blue open symbols: various randomly chosen orientations. The rates are scaled by $1/\varepsilon_{\tm{r}}$ with $\varepsilon_{\tm{r}} = 3$. Shaded gray area: $r^{-6}$ dependence. Shaded pink area: $r^{-8}$ dependence. Dashed line: extrapolation of $r^{-6}$ dependence to lower $r$.}
    \label{fig:Rate_to_distance}
\end{figure}

\textbf{Figure~\ref{fig:Rate_to_distance}} shows the calculated TPQ rate for a pair of {\emit} and {\hostc} molecules, in the relaxed structure of a COSMO embedding, as a function of the centre-of-mass distance. The rates were calculated from Fermi's Golden rule (Equation~\ref{eq:FermiGoldenRuleTPQ} in \hyperref[sec:Methods]{Methods}), including vibrations and using exactly calculated direct Coulomb couplings between the initial and final excitonic states. 
The connected data points show results for a fixed, randomly chosen, mutual orientation of the pair of molecules. 
The figure also shows results for the total TPQ rate for randomly oriented pairs at various distances (open symbols), where a wide distribution ($\sim 3$ orders of magnitude) in the results can be observed for each distance. The chosen single fixed mutual orientation happens to be such that the resulting TPQ rate lies on the high side in the distribution.

As expected, for large distance the TPQ rate follows a Förster-type $r^{-6}$ dependence (gray band). 
We observe that for $r \lesssim 5$ nm the TPQ rate starts to exceed the Förster rate: the data points lie above the dashed line extrapolated from the Förster regime. For smaller distances the rate is well described by an $r^{-8}$ dependence (pink band), which is indicative of a dipole-quadrupole coupling. For even smaller distances, approaching 1 nm, the distance dependence deviates from the $r^{-8}$ dependence. At such small distances, the donor and acceptor NTOs start to overlap, so that a description in terms of multipole couplings is not useful. We note that a dominance of dipole-quadrupole coupling over dipole-dipole coupling in intermolecular energy transfer may occur for certain mutual orientations of two molecules for which the dipole-dipole coupling is small or even disappears \cite{sissaForsterFormulationResonance2011}. This is not the case for the specific mutual orientation in Fig.~\ref{fig:Rate_to_distance}, where, in contrast, the dipole-dipole coupling is stronger than on average.

Equivalently to the dipolar molar absorption coefficient, one can define a \textit{quadrupolar} molar absorption coefficient of the acceptor A as 
\begin{equation}
     \epsilon^{\tm{q}}_{\tm{A}} \of{E} \equiv \frac{\pi N_{\tm{A}} E^3 n^3}{15 \epsilon_0 \epsilon_{\tm{r}} \ln \of{10} \hbar^3 c^3} \sum_{j} Q_{j}^2 \rho^{\tm{FC}}_{j} \of{E}.
\end{equation}
Here, $\rho_{\tm{FC},j} \of{E}$ is the Franck-Condon weighted density of states and $Q_{j}^2$ denotes the square of the transition quadrupole moment of excitation $j$ on the acceptor molecule, which can be defined as \cite{buckinghamMolecularQuadrupoleMoments1959}

\begin{align}\label{eq:SquareQuadrupoleMoment}
    Q_{j}^2 \equiv & \frac{3}{2}\sum_{\alpha\beta} Q_{j,\alpha\beta}^2 - \frac{1}{2} \of{\sum_{\alpha} Q_{j,\alpha\alpha}}^2,
\end{align}
where $Q_{j,\alpha \beta} = \bra{0} Q_{\alpha \beta} \ket{j}$ ($\ket{0}$ is the ground state and $\ket{j}$ is the state of excitation $j$). The quadrupole moment operator is $Q_{\alpha \beta} = -e \sum_{m} r_{m, \alpha} r_{m,\beta}$, where the sum over $m$ runs over all electrons. We make the choice to take the centre-of-mass position of the molecule as the origin and to define the Cartesian components $r_{m,\alpha}$ relative to this origin. In the case of a non-vanishing dipole moment, the choice of the origin has an influence on the transition quadrupole moment  \cite{bernadotteOriginindependentCalculationQuadrupole2012}. Taking higher transition multipole moments into account will remove the origin dependence. The $r^{-8}$ dependence found in Figure~\ref{fig:Rate_to_distance} suggests that our choice of the centre of mass as the origin is appropriate, at least down to the distance where this dependence fits the data.

The red line in Figure~\ref{fig:ExperimentalSpectraOverlap}(b) shows the quadrupolar molar absorption coefficient $\epsilon^{\tm{q}}_{\tm{A}} \of{E}$ of {\hostc}, obtained by averaging over the spectra of $21$ randomly chosen conformers in the thin film slice of Figure~\ref{fig:Figuur_1_workflowv2}(b) (Supplemental Figure~\ref{fig:IndividualQuadrupoleAbsorptionSpectra}  \cite{SupplementaryInformation} shows the absorption spectra of the individual conformers). The conformers were put in a COSMO embedding with $\varepsilon_{\tm{r}} = 3$. This simplified approach as compared to the calculation of the dipolar molar absorption coefficient in the thin film is justified by the small difference in the dipolar absorption spectra between solution (green line in Figure~\ref{fig:ExperimentalSpectraOverlap}(a)) and \mbox{thin film} (blue line in Figure~\ref{fig:ExperimentalSpectraOverlap}(b)). Because of computational restrictions, we chose for this simplification and for a smaller amount of conformers in the quadrupolar than in the dipolar calculations.

The main peaks in the quadrupolar absorption spectrum lie approximately at the same energies as in the dipolar spectrum. The peak around 1 eV in the quadrupolar absorption spectrum is somewhat blueshifted with respect to the dipolar absorption spectrum. This is a result of slightly upward shifted excitation energies caused by the different modeling of the embedding. The blueshift in the peak around 3 eV is not a result of shifted energies, but of different magnitudes of the transition quadrupole and dipole moments of the excitations. Both the dipolar and quadrupolar absorption spectra show a shoulder around $2.6$ eV, which partially overlaps with the emission spectrum of {\emit}. The shoulder is relatively more pronounced in the quadrupolar absorption spectrum, because of the relatively large transition quadrupole moments of the concerned excitations.

It is clear from Figure~\ref{fig:ExperimentalSpectraOverlap}(b) that the quadrupolar absorption spectrum of {\hostc} is much smaller than the dipolar absorption spectrum (note the multiplication of the quadrupolar spectrum by a factor $10^4$). This expresses the fact that in an ordinary absorption experiment, involving electromagnetic waves that are generated by oscillators at a far distance, it is sufficient to only consider dipolar absorption. However, as shown in Figure~\ref{fig:Rate_to_distance}, quadrupolar absorption is not negligible for electromagnetic waves generated by oscillators at small distances, even becoming dominant at distances of a few nm.

Analogously to the dipole-dipole rate in Equation~\ref{eq:BasicFoersterFormula}, the dipole-quadrupole rate for TPQ is
\begin{align}\label{eq:Rate_dipole_quadrupole}
    k_{\tm{dq}} \of{r} =& \frac{45 c^6 \hbar^6 \ln \of{10} \eta \expec{\kappa^2_{\tm{dq}}}}{8 \pi n^6 N_{\tm{A}} \tau r^8}  \int_{0}^{\infty}  \frac{I_{\tm{D}} \of{E} \epsilon^{\tm{q}}_{\tm{A}} \of{E}}{E^{6}} \diff E \nonumber \\
    \equiv& \frac{1}{\tau_{\tm{r}}} \of{\frac{R_{\tm{dq}}}{r}}^8.
\end{align}
 For the expectation value of the square of the dipole-quadrupole orientation factor we take again the dynamic isotropic averaging value $\expec{\kappa^2_{\tm{dq}}}=1$ (see Section \ref{subsec:dipquadrate} in \hyperref[sec:Methods]{Methods}). Applying Equation~\ref{eq:Rate_dipole_quadrupole} to the spectra in Figure~\ref{fig:ExperimentalSpectraOverlap}(b) yields $R_{\tm{dq}} = 2.6$ nm for the TPQ dipole-quadrupole radius. Since the theoretical TPQ Förster (dipole-dipole) radius was found to be $2.7$ nm, this implies that below a distance of \mbox{$R_{\tm{qd}}^4/R_{\tm{F}}^3 \approx 2.3$ nm} the dipole-quadrupole contribution to the TPQ rate dominates over the dipole-dipole contribution, as is the case in Figure~\ref{fig:Rate_to_distance}. Since the average distance at which a TPQ event occurs was estimated to be $\expec{r} \approx 1.5$ nm, we can conclude that in our system TPQ caused by dipole-quadrupole coupling dominates over TPQ caused by \mbox{dipole-dipole} coupling.

In the modeling of the TPQ experiments in Ref.~\cite{ligthartMechanisticDescriptionEfficiency2021}, yielding $R_{\tm{F}} = 3.8$ nm, the TPQ process was assumed to be a Förster process. In view of our conclusion that quadrupole-dipole TPQ is more important than dipole-dipole TPQ, the experimentally derived Förster radius should be considered as an effective Förster radius, accounting for all contributions to the TPQ rate. Because polaron hopping is much faster than exciton decay, we can neglect correlations between positions of polarons and excitons \cite{taherpourEffectsCorrelationsTriplet2024}.

This means that we can assume that the occupational probability of host molecules by a polaron in the neighborhood of an exciton on an emitter molecule that has not yet decayed or undergone TPQ does not depend on the distance between the polaron and the exciton. This allows us to define an effective Förster radius as
\begin{align}
    R_{\tm{F,eff}} \equiv \of{R^6_{\tm{F}} + c_{\tm{dq}} \frac{R^8_{\tm{dq}}}{r_0^2}}^{1/6},\label{eq:RFeff}
\end{align}
where $c_{\tm{dq}}$ is a numerical factor of order 1, which depends on the particular ordering of the host molecules around a emitter molecule. Modeling the ordering as a continuum of host molecules beyond a nearest-neighbor distance $r_0\approx 1$ nm, as we did in Equation~\ref{eq:TypicalForsterInteractionLength}, yields $c_{\tm{dq}} = 5/7$, leading to $R_{\tm{F,eff}} \approx 3.5$ or 3.7 nm using the SO method, depending on whether we take the theoretical value \mbox{$R_{\tm{F}} = 2.7$ nm} or the experimental value \mbox{$R_{\tm{F}} = 3.1$ nm}, respectively. Taking as a more realistic ordering a simple cubic lattice, we find $c_{\tm{dq}} \approx 0.83$ (see Section \ref{subsec:Numericallattefactors} in \hyperref[sec:Methods]{Methods}), leading to $R_{\tm{F,eff}} \approx 3.6$ or $3.7$ nm. The closeness of the continuum and lattice values indicates that the precise morphology of the host molecules is not very important. All values are reasonably close to the value $R_{\tm{F}} = 3.8$ nm found in Ref.~\cite{ligthartMechanisticDescriptionEfficiency2021}. We conclude from this that taking into account the dipole-quadrupole coupling contribution to TPQ can, at least for a large part, explain the much higher value of $R_{\tm{F}}$ obtained from the modeling of TPQ experiments on an {\host}:{\emit} device (FF method) than from the overlap between the emission spectrum of {\emit} and the absorption spectrum of {\hostc} (SO method).

We limited in the present work our analysis to the consideration of \mbox{dipole-quadrupole} coupling and did not consider higher-order multipole couplings. Our exact direct Coulomb coupling results in Figure~\ref{fig:Rate_to_distance} do not show a region that can be described by an $r^{-2n}$ dependence with $n>4$, which would be indicative of the dominance of a higher-order multipole coupling than dipole-quadrupole coupling. For very small distances, the multipole expansion loses its usefulness and only calculations where the direct Coulomb integrals are exactly calculated yield reliable results. 

We should stress that in our calculation of the TPQ rate we have only accounted for direct Coulombic coupling between an exciton on an {\emit} molecule and a polaron on an {\hostc} molecule (see Figure~\ref{fig:Rate_to_distance}). We did not account for TPQ mediated by exchange coupling, where the exciton on an {\emit} molecule is transferred to an {\hostc} molecule because of wavefunction overlap. TPQ mediated by exchange coupling could become important for very small intermolecular distances. Because of the exponential decay of the wavefunctions, a calculation of exchange coupling-mediated TPQ requires a detailed analysis of the morphological aspects of the embedding of {\emit} into {\host} and a very precise calculation of the concerned wavefunctions. We remark that, even if TPQ mediated by exchange coupling will turn out be important, this will not invalidate our conclusion that in the TPQ process considered in the present work \mbox{dipole-quadrupole} coupling is more important than Förster-type dipole-dipole coupling.

\section{\label{sec:Discussion}Conclusion and Outlook}
The present case study of TPQ in the system {\host}:{\emit} has implications that are relevant to a broad range of phosphorescent emitter-host systems. Iridium-based phosphorescent emitters are generally rather compact, so that in the Coulombic coupling with excitations of charged host molecules the triplet excitons on these emitters can, even at short distances, be approximately described in the dipole approximation. On the other hand, various commonly used host molecules are often rather extended and flexible, like {\host}, so that the Coulombic coupling between excitations on a host molecule and an Ir-based emitter molecule can at short distances not be properly described by the Förster dipole-dipole approximation (as illustrated in Figure~\ref{fig:Figuur_2_mMTDATA_Irppy2acac}). When the polarons are confined to the host and hop quickly among the host molecules, the polarons can closely approach triplet excitons on emitter molecules within the lifetime of the excitons, so that short distances between triplet excitons and polarons will regularly occur. An important consequence is that it is not sufficient to only consider the overlap of the emission spectrum of the emitter and the dipolar absorption spectrum of the charged host in order to judge the importance of TPQ (see Figure~\ref{fig:ExperimentalSpectraOverlap}(a)). The present work shows the importance of considering also the quadrupolar absorption spectrum of the charged host (see Figure~\ref{fig:ExperimentalSpectraOverlap}(b)).

We conclude that in the class of cases that TPQ is governed by the transfer of the excitation energy of a triplet on an Ir-based emitter to a polaron that is confined to the host, a description of TPQ in terms of Förster \mbox{dipole-dipole} transfer could fail, because of three reasons: (1) the extendedness of the host molecule, allowing excitations with intramolecular charge transfer over a large distance, (2) the short distance at which a polaron on the host can approach the triplet on the emitter, and (3) the symmetry of the excitonic wavefunctions on the charged host, which could allow quadrupolar but forbid dipolar couplings. On the other hand, we expect that in the class of cases that TPQ is governed by the transfer of the excitation energy of a triplet exciton on an Ir-based emitter to a polaron that is also on the emitter (confinement of the polarons to the emitter), a description of TPQ in terms of Förster transfer is very likely sufficient, because of the compactness of the emitter molecule and the resulting limited distance of charge transfer involved in its excitations and because of the large average distance between emitter molecules. It would be of great interest to investigate these expectations in combined experimental and theoretical studies in order to find combinations of emitters and hosts with reduced TPQ.

In the case that TPQ would have been analyzed with the SO method, the TPQ rate would have been significantly underestimated and incorrect conclusions about operational lifetimes and efficiencies been made. It is important to be able to determine correct TPQ rates in order to find stable blue OLEDs. Discrepancies between the spectral overlap (SO) method and methods comparable to the Förster-fit (FF) method to obtain the Förster radius for excitonic processes have also been reported for other cases, such as exciton diffusion in biological systems \cite{colbowEnergyTransferPhotosynthesis1973,fennelForstermediatedSpectralDiffusion2012,vuojolaDistanceTemperatureDependency2011,wareExperimentalStudyEnergy1961,stryerEnergyTransferSpectroscopic1967a,rollerDETERMINATIONFORSTERCRITICAL2004} and in quantum dots \cite{morkMagnitudeForsterRadius2014a}. Like in the present work, in nearly all cases the Förster radius obtained from the SO method is smaller than determined with FF-type methods, suggesting a systematic trend. Therefore, the conclusion from our work that considering the Förster \mbox{dipole-dipole} approximation is insufficient to describe TPQ in the studied system and that dipole-quadrupole coupling needs to be considered may be applicable to excitonic processes in other systems as well. 
We finally note that dipole-quadrupole couplings have also been shown to be important in completely different contexts, such as van der Waals forces \cite{margenauRoleQuadrupoleForces1931,mayerDispersionPolarizabilityVan1933,tangUpperLowerBounds1976} and nuclear magnetic resonance \cite{hartmannNuclearDoubleResonance1962}, so that opportunities for cross-fertilization could be present.

\section{\label{sec:Methods}Methods}

\subsection{Solution workflow}\label{subsec:solutionfilmworkflow}
First, a molecular dynamics (MD) simulation is performed using DFT-calculated gradients employing the PBE \cite{perdewGeneralizedGradientApproximation1996} exchange-correlation (XC) functional with DZP basis set \cite{vanlentheOptimizedSlatertypeBasis2003} for the positively charged state of {\host}, {\hostc}, in a COSMO \cite{pyeImplementationConductorlikeScreening1999} embedding with $\varepsilon_{\tm{r}}=3$ and a solvent radius of $3$ \r{A}, as implemented in the ADF program of the AMS software suite \cite{teveldeChemistryADF2001}. The molecule is attached to a chain of Nosé–Hoover thermostats \cite{kleinermanImplementationsNoseHoover2008} with a temperature of $300$ K and a damping constant of $50$ fs. All bonds involving hydrogen atoms are constrained, allowing for MD time steps of $1$ fs. The molecular geometry is recorded at $0.1$ ps time intervals over a total simulation time of $10$ ps. The initial equilibration period of about $0.5$ ps is disregarded (see Figure \ref{fig:Figuur_1_workflowv2}), yielding an ensemble of $95$ conformers for which the absorption spectrum is determined.

The absorption spectrum of the {\host} conformers in their charged state, {\hostc}, is determined from the energies and transition dipole moments of the $25$ energetically lowest excitations, calculated with \mbox{time-dependent} density functional theory (TD-DFT) \cite{vangisbergenImplementationTimedependentDensity1999,wangExcitationEnergiesD12004}. All TD-DFT calculations in this work use the \mbox{Tamm-Dancoff} approximation \cite{hirataTimedependentDensityFunctional1999} (TDA). These calculations are performed using the B3LYP XC functional \cite{stephensInitioCalculationVibrational1994}, the TZ2P basis set \cite{vanlentheOptimizedSlatertypeBasis2003}, and scalar relativistic spin-orbit coupling corrections within the zeroth-order regular approximation (ZORA) \cite{vanlentheRelativisticTotalEnergy1994} to the Dirac equation, as implemented in the ADF program. Also in these calculations the COSMO embedding is used.

\subsection{Thin film workflow}\label{subsec:thinfilmworkflow}
The OLED deposition and properties workflow available in the AMS software suite \cite{rugerSCMOLEDDeposition} is used for determining the absorption spectra of conformers in the thin film. In the MD simulation of the deposition of the {\host} thin film, first the atomic partial charges are calculated for the {\host} molecule in vacuum using the DZP basis set and the PBE XC functional, with multipole derived charges evaluated at the quadrupole level (MDC-q) \cite{swartChargeAnalysisDerived2001}. In the MD simulation, every $1$ ps an {\host} molecule is launched at a $6$ by $6$ nm square sheet of graphene in the $x$-$y$ plane. The forces are calculated using the UFF force field \cite{rappeUFFFullPeriodic1992,addicoatExtensionUniversalForce2014,coupryExtensionUniversalForce2016}, which includes van der Waals forces, with the Coulomb forces between the atomic partial charges of different molecules added. The positions of the atoms and their velocities are updated with a time step of $1$ ps. All bonds with hydrogen atoms are constrained in length. The velocities of the atoms are coupled to a Berendsen thermostat with a temperature of $600$ K and a damping constant of $100$ fs. The simulations are continued until $315$ molecules are deposited, after which the graphene layer is removed. 
The system is then additionally coupled for $450$ ps to a Martyna-Tobias-Klein (MTK) barostat \cite{martynaConstantPressureMolecular1994a} with a pressure of $1$ atm and a damping constant of $1$ ps. After performing this simulation for $300$ ps, the temperature of the thermostat is ramped down linearly to $300$ K over the course of $100$ ps to bring the film to ambient conditions. Lastly, a geometry optimization is performed with the barostat at 1 atm, where the lattice vectors in the $x$-$y$ plane are optimized. This finally leads to a box with a lateral size of approximately $5.5$ by $5.5$ nm containing the thin film with a thickness of approximately $14.7$ nm.

Next, the absorption spectra are determined of \mbox{163 conformers} in their charged state, randomly chosen from the inner 5.5 nm thick slice of the simulated film. The geometry of each charged {\hostc} conformer is optimized in its environment, which is kept fixed. The environment is taken to consist of surrounding molecules that have a nearest atom-atom distance below 0.5 nm with atoms of the considered {\hostc} conformer, excluding hydrogen atoms. The interaction of the charged conformer with the surrounding molecules is modeled by the DRF QM/MM embedding model \cite{jensenDiscreteSolventReaction2003}, using the calculated partial atomic charges of all the atoms of the surrounding molecules. 
After the geometry optimization of each charged {\hostc} conformer in its fixed environment, a single-point TD-DFT calculation is performed of the charged conformer in its environment to determine the energies and transition dipole moments of the $25$ energetically lowest-lying excitations. In both the geometry optimization and the single-point TD-DFT calculation use is made of the TZ2P basis set and B3LYP functional, with spin-orbit coupling taken into account in the ZORA.

\subsection{Calculation of quadrupolar absorption spectra}\label{subsec:quadrupoleworkflow}

The quadrupolar absorption spectra of 21 host molecules in their charged state randomly chosen from the inner 5.5 nm thick slice of the simulated thin film are calculated with the PySCF software \cite{sunPySCFPythonbasedSimulations2018,sunRecentDevelopmentsSCF2020}. The {\hostc} conformers are placed in a COSMO embedding with \mbox{$\varepsilon_{\rm r} = 3$}. The energies and transition quadrupole moments of the $25$ lowest-energy excitations of each charged conformer are obtained with TD-DFT, using the B3LYP XC functional and the def2-TZVP \cite{hellwegDevelopmentNewAuxiliary2014} basis set.

\subsection{Calculation of TPQ rates with exact direct Coulomb integral coupling}

To calculate the TPQ rates with exact direct Coulomb integral coupling, ground-state vacuum conformers of both an {\emit} and {\hostc} molecule are placed in a COSMO environment with \mbox{$\varepsilon_{\rm r} = 3$}. Various \mbox{centre-of-mass} distances are chosen and fixed (closed symbols in Figure \ref{fig:Rate_to_distance}) or random (open symbols) mutual orientations of the molecules. The wavefunctions of the $25$ lowest-energy excitations are calculated for both {\emit} and {\hostc} with TD-DFT using the B3LYP functional and the def2-SVP \cite{hellwegDevelopmentNewAuxiliary2014} basis set, as implemented in the PySCF software. The direct Coulomb integrals are then exactly calculated for the $6$ lowest-energy singlet and $18$ lowest-energy triplet excitations of {\emit} and excitations $3$ to $10$ of {\hostc}. Only the inclusion of these excitations is finally necessary in the calculation of the TPQ rate. Spin-orbit coupling is included in a perturbative way \cite{wangSimplifiedRelativisticTimedependent2005} in the ZORA and the spin-orbit coupling matrices between excitations are calculated with ADF.

\subsection{Calculation of Franck-Condon weighted density of states}
The Franck-Condon (FC) weighted density of states (DOS) of excitations is calculated within the vertical gradient approximation to the displaced oscillator model \cite{desouzaTheoreticalPredictionFluorescence2018}, as in Ref.~\cite{devriesHighEnergyAcceptor2020}. Within this approximation the frequencies $\omega_m$ of each phonon mode $m$ are calculated from the mass-weighted Hessian of the ground state conformation in the ground state electronic configuration. The displacements are calculated with a single Newtonian update step using the ground state Hessian and excited state gradients obtained within TD-DFT \cite{sethTimedependentDensityFunctional2011a}, again using the B3LYP XC functional and TZ2P basis set, which enables the calculation of the reorganization energies $\lambda_{im}$ of excitation $i$ associated with phonon mode $m$  \cite{devriesFullQuantumTreatment2018}. The FC weighted density of states can then be calculated as \cite{devriesFullQuantumTreatment2018} 
\begin{align}\label{eq:FCWD}
    \rho^{\tm{FC}}_{i} \of{\Delta E_i} = \frac{1}{2 \pi \hbar} \int_{- \infty}^{+\infty} e^{ \frac{i \of{\Delta E_i - \lambda_{\tm{cl}}} t}{\hbar} - \frac{\lambda_{\tm{cl}} k_{\tm{B}} T t^2}{\hbar^2} } I_{i} \of{t} d t,
\end{align}
where $\lambda_{\tm{cl}}$ is the ``classical'' reorganization energy of low-frequency phonon modes that are not explicitly taken account, where we take $\lambda_{\tm{cl}} = 0.05$ eV, and $I_{i}$ accounts for the contributions from high-frequency phonon modes \cite{devriesFullQuantumTreatment2018},
\begin{align}
    I_{i} \of{t} = e^{\sum_m 2 \frac{\lambda_{im}}{\hbar \omega_m} \of{\coth \of{\frac{\hbar \omega_m}{2 k_{\tm{B}} T }} \off{\cos \of{\omega_m t} -1} + i \sin \of{\omega_m t} }}.
\end{align}
In Equation~\ref{eq:FCWD} $\Delta E_i$ is the difference in energy between the initial and final states. For donors we have $\Delta E_i = E_i - E$ and for acceptors we have $\Delta E_i = E - E_i$, where $E$ is the energy of the emitted or absorbed photon and $E_i$ is the energy of excitation $i$. 

In the case that we have both a donor and an acceptor, their Franck-Condon weighted density of states should be combined to form the total Franck-Condon weighted density of states \cite{devriesHighEnergyAcceptor2020}, 
\begin{align}
    \rho^{\tm{FC}}_{ij} \of{E_j - E_i} = \int^{+\infty}_{- \infty} \rho^{\tm{FC}}_{i} \of{E_i - E_j - E} \rho^{\tm{FC}}_{j} \of{E} d E,
\end{align}
where the excitations $i$ and $j$ are located on the donor and acceptor, respectively and $\Delta E_{ji} = E_j - E_i$.

\subsection{Calculation of the dipolar absorption spectrum}
The molar absorption coefficient of an acceptor (A) can be calculated as \cite{devriesHighEnergyAcceptor2020} 
\begin{align}
    \epsilon_{\tm{A}} \of{E} = \frac{\pi N_{\tm{A}} n E}{3 \varepsilon_0 \varepsilon_{\tm{r}} \hbar^2 c \ln \of{10}} \sum_{j} \mu^2_j \rho^{\tm{FC}}_{j} \of{E}, 
\end{align}
where $\mu^2_{j}$ denotes the square of the transition dipole moment of the $j$th excitation of the acceptor.

The emission intensity of a donor (D) can be determined as \cite{desouzaTheoreticalPredictionFluorescence2018}
\begin{equation}
    i_{\tm{D}}\of{E} = \frac{n E^3 }{3 \pi \varepsilon_0 c^3 \hbar^4} \sum_i p_i \mu^2_{i} \rho^{\tm{FC}}_{i} \of{E},
\end{equation}
where $p_i$ is the probability that the donor is initially in excitation $i$, for which we take the Boltzmann probability at $300$ K. The normalized emission spectrum can then be calculated
\begin{equation}
    I_{\tm{D}} \of{E} = \frac{i_{\tm{D}}\of{E}}{\int^{\infty}_{0} i_{\tm{D}}\of{E} d E} = \tau_{\tm{r}} i_{\tm{D}}\of{E}.
\end{equation}

\subsection{Calculation of the TPQ rates with Fermi's Golden Rule}

All calculations of TPQ rates in this work are based on Fermi's Golden Rule,
\begin{align}\label{eq:FermiGoldenRuleTPQ}
    k = \frac{2 \pi}{\hbar} \sum_{ij} p_i \abss{ \bra{j}  \mathcal{H}' \ket{i}}^2 \rho^{\tm{FC}}_{ij} \of{E_j - E_i},
\end{align}
where $\mathcal{H}'$ is the perturbing Hamiltonian, $E_i$ and $E_j$ are the energies of the system in the initial and final states, $\ket{i}$ and $\ket{j}$, respectively. The three energetically lowest-lying near-degenerate triplet excited states of the emitter {\emit}, combined with the, doubly degenerate, {\hostc} ground state, are considered as possible initial states. The final states are the doublet excited states of {\hostc} combined with the ground state of {\emit}. 
The wavefunctions of the excitations are obtained from TD-DFT with the TDA and are expressed as linear combinations of combined occupied and empty orbitals \cite{wehnerIntermolecularSingletTriplet2017}, $\ket{j} = \sum_{\alpha a} A_{\alpha a,j} \ket{\alpha a}$ and $\ket{i} = \sum_{\beta b} B_{\beta b,i} \ket{\beta b}$, where $\alpha$, $\beta$ run over occupied orbitals and $a$, $b$ over unoccupied orbitals. The wavefunctions of the excitations are orthogonalized between acceptor and emitter, as described in Ref. \cite{wehnerIntermolecularSingletTriplet2017}. 

We consider for $\mathcal{H}'$ the Coulomb coupling between electrons, such that \mbox{$\mathcal{H}' = \sum_{mn} \frac{e^2}{4 \pi \varepsilon_{0} \varepsilon_{\tm{r}}} \frac{1}{\abss{\vect{r}_m - \vect{r}_n}}$}, where $m$ runs over all electrons on the donor molecule and $n$ over all electrons on the acceptor molecule. 
With the orthogonalized wavefunctions of the excitations, the rate for TPQ becomes
\begin{align}
    k = & \frac{2 \pi}{\hbar} \of{\frac{e^2}{4 \pi \varepsilon_{0} \varepsilon_{\tm{r}}}}^2 \sum_{ij} p_i \abss{J_{ij}}^2 \rho^{\tm{FC}}_{ij} \of{E_j - E_i}.
\end{align}
We do not consider the exchange part, such that $J_{ij} = \sum_{\alpha \beta a b} A^*_{\alpha a,j} B_{\beta b,i} \of{\alpha a|\beta b}$ is the direct Coulomb coupling between the triplet exciton $i$ on {\emit} and the doublet excitation $j$ on {\hostc}, and $\of{\alpha a|\beta b}$ is the two-electron Coulomb integral in the chemist's notation. 

\subsection{Calculation of the dipole-quadrupole coupling}\label{subsec:dipquadrate}
The direct Coulomb coupling $J_{ij}$ can be expanded as a series in the centre-of-mass distance $r$. The terms containing monopoles can be ignored, as the considered excitations are charge neutral. In that case the first two non-zero terms of the series are
\begin{widetext}
\begin{align}
    J_{ij} \approx & \frac{1}{r^3} \of{\sum_{k} \mu_{i,k} \mu_{j,k} - 3 \sum_{k l} \mu_{i,k} \mu_{j,l} \widehat{r}_{k} \widehat{r}_{l} }\nonumber\\
        & + \frac{3}{2 r^4} \ofl{\sum_{k l} \off{{\mu}_{i,k} {Q}_{j,l l} \widehat{r}_{k} + 2 {\mu}_{i,k} {Q}_{j,k l} \widehat{r}_{l}}} \ofr{-5 \sum_{k l m} {\mu}_{i,k} {Q}_{j,l m} \widehat{r}_{k} \widehat{r}_{l} \widehat{r}_{ m}}\nonumber\\
        = & \frac{{{\mu}_i} {{\mu}_j} }{r^3} \of{\sum_{k} \widehat{\mu}_{i,k} \widehat{\mu}_{j,k} - 3 \sum_{k l} \widehat{\mu}_{i,k} \widehat{\mu}_{j,l} \widehat{r}_{k} \widehat{r}_{l} }\nonumber\\
        & + \frac{3 {{\mu}_i} {{Q}_j}}{2 r^4} \ofl{ 5 \sum_{k l m} \widehat{\mu}_{i,k}  \widehat{Q}_{j,l m} \widehat{r}_{k} \widehat{r}_{l} \widehat{r}_{ m}} - \ofr{\sum_{k l} \off{\widehat{\mu}_{i,k} \widehat{Q}_{j,l l} \widehat{r}_{k} + 2 \widehat{\mu}_{i,k} \widehat{Q}_{j,k l} \widehat{r}_{l}}}\nonumber\\
        = & \frac{{{\mu}_i} {{\mu}_j} \kappa_{ij}}{r^3} + \frac{3 {{\mu_i}} {{Q}_j} \kappa_{\tm{dq},ij}}{2 r^4},
\end{align}
\end{widetext}
where $\mu_i$ ($\mu_j$) and ${Q}_i$ (${Q}_j$) are the magnitudes of the transition dipole and quadrupole moments, respectively, of the excitation $i$ ($j$) on the donor (acceptor). The additional indices $k,l,m$ run over the three Cartesian components $\offf{x,y,z}$. The vector $\vect{r}$ is the centre-of-mass distance vector between donor and acceptor. The hats indicate vectors or tensors of unit size. We note that we assume that only excitation $j$ on the acceptor has a transition quadrupole moment. Taking the isotropic average over the square of the dipole-dipole and dipole-quadrupole orientation factors yields $\expec{\kappa^2}=2/3$ and $\expec{\kappa^2_{\tm{dq}}}=1$, respectively, where we can drop the indices $i$ and $j$, because the isotropic average is independent of $i$ and $j$.

\subsection{Calculation of numerical lattice factor $c_{\tm{dq}}$}\label{subsec:Numericallattefactors}
The numerical factor in Equation \ref{eq:RFeff} for calculating the effective Förster radius in the presence of dipole-dipole and dipole-quadrupole coupling can be obtained by taking the ratio of two lattices sums,
\begin{align}
    c_{\tm{dq}} = a \frac{\sum_{p \in \tm{lattice}} r_p^{-4} }{ \sum_{p \in \tm{lattice}} r_p^{-3} },
\end{align}
where $a$ is the lattice constant and where the origin should be excluded in each sum. Evaluating the sums for a simple cubic lattice yields $c_{\tm{dq}} = 0.8267...$.

\section{Acknowledgements}
This work is part of the project ``Suppressing Exciton Quenching in OLEDs: an Integrated Approach" (SEQUOIA), with Project No. 18975, of the research program ``Open Technology” of the Dutch Research Council, Nederlandse Organisatie voor Wetenschappelijk Onderzoek (NWO). The project is jointly financed by NWO, Merck KGaA, SCM B.V., and Simbeyond B.V. The authors thank Dr.\ Erik van Lenthe, Dr.\ Robert Rüger, MSc.\ Bart Klumpers, Dr.\ Franco Egidi, and Dr.\ Stan van Gisbergen from Software for Chemistry \& Materials B.V. (SCM), MSc.\ Chima Chibueze and Prof. Lucas Visscher from the Vrije Universiteit Amsterdam, MSc.\ Hiroki Tomita, MSc.\ Christian McDonald, MSc.\ Lois Fernandez Miguez, dr.\ Engin Torun, and Dr.\ Christ Weijtens from Eindhoven University of Technology, and Dr.\ Philipp Sch\"{u}tz, Dr.\ Alexander Schubert, Dr.\ Falk May, and Dr.\ Christof Pflumm from Merck KGaA for discussions.

The data that support the findings of this article are openly available \cite{Hoesel2025DataSetThisPaper}.

This paper has been published in Physical Review B \cite{vanhoeselDipolequadrupoleCouplingTriplet2025}.

\bibliography{mainbib}

\setcounter{figure}{0}  
\renewcommand{\thefigure}{S\arabic{figure}}
\setcounter{equation}{0}  
\renewcommand{\theequation}{S\arabic{equation}}
\setcounter{section}{0}  
\renewcommand{\thesection}{S\roman{section}}

\section{Spectra of individual conformers}
\begin{figure}[!h]
    \centering
    \includegraphics[width=8.5cm]{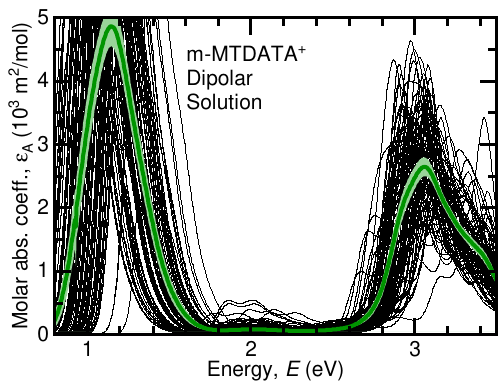}
    \caption{Black lines: calculated molar absorption spectra of each of the $95$ {\hostc} conformers from the solution simulation. Green line: average of the individual spectra. Shaded band: $2 \sigma$ confidence interval of the average.}
    \label{fig:IndividualDipoleAbsorptionSpectraSolution}
\end{figure}

\begin{figure}[!h]
    \centering
    \includegraphics[width=8.5cm]{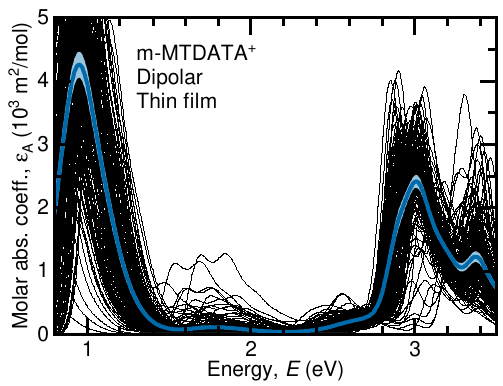}
    \caption{Black lines: calculated molar absorption spectra of each of the $163$ {\hostc} conformers from the thin film simulation. Blue line: average of the individual spectra. Shaded band: $2 \sigma$ confidence interval of the average.}
    \label{fig:IndividualDipoleAbsorptionSpectraMorphology}
\end{figure}

\begin{figure}[!h]
    \centering
    \includegraphics[width=8.5cm]{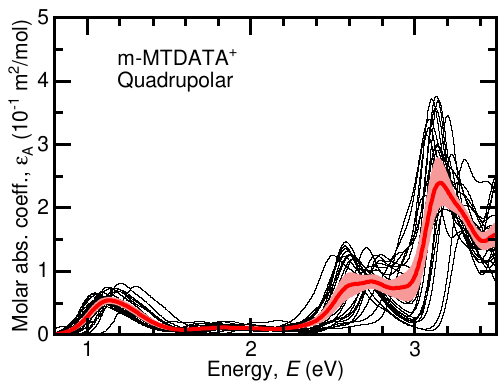}
    \caption{Black lines: calculated quadrupolar molar absorption spectra of $21$ {\hostc} conformers from the thin film simulation. Red line: average of the individual spectra. Shaded band: $2 \sigma$ confidence interval of the average.}
    \label{fig:IndividualQuadrupoleAbsorptionSpectra}
\end{figure}

Figures \ref{fig:IndividualDipoleAbsorptionSpectraSolution}, \ref{fig:IndividualDipoleAbsorptionSpectraMorphology} and \ref{fig:IndividualQuadrupoleAbsorptionSpectra} show the spectra as a function of photon energy of the individual conformers considered for the averaged spectra.

\newpage
\section{Self-absorption}\label{sec:SelfAbsorption}
The calculation of the Förster radius is sensitive to the shape of the emission spectrum of the emitter, \emit. The measured emission spectrum might be influenced by self-absorption of emitted photons by the emission layer itself, whereas the spectrum without self-absorption should be used for determining the TPQ Förster radius. How the shape of the emission spectrum is precisely changed by self-absorption is a difficult question, because self-absorption of a photon in a film may lead to re-emission of a photon by the excited molecule at another wavelength and in another direction. We will for simplicity assume that the {\emit} molecules are uniformly distributed throughout the layer and that re-emission does not occur. According to the Beer-Lambert law the observed emission spectrum $I_{\text{obs}}$ is under these assumptions related to the actual emission spectrum $I$ by
\begin{align}
    I_{\text{obs}} \of{E} = & I \of{E} \int^{d}_{0} \frac{e^{- \epsilon \of{E} c_{\text{mol}} x}}{d} \diff x  \nonumber\\
    = & I \of{E}  \frac{1 - e^{- \epsilon\of{E} \, c_{\text{mol}} d}}{\epsilon\of{E} \, c_{\text{mol}} d}  \nonumber\\
    = & I \of{E} \frac{- \alpha \of{E}  }{ \ln \of{ 1 - \alpha \of{E} } } \nonumber\\
    \geq &  I \of{E} \of{1 - \alpha \of{E} } ,
\end{align}
where $d$ is the thickness of the thin film ($50$ nm in the experiment of Ref.~\cite{ligthartEffectTripletConfinement2018}), $\epsilon \of{E}$ is the molar absorption coefficient of the thin film and $\alpha \of{E} = 1 - e^{- \epsilon \of{E} \, c_{\text{mol}} d}$ denotes the absorption fraction of the film. For the molar concentration $c_{\text{mol}}$ we take $1660$ mol/m$^3$, corresponding to 1 molecule per nm$^3$. We can rewrite the above expression as 
\begin{align}\label{eq:CorrectedEmissionSpectrum}
    I \of{E} = \frac{- \ln \of{ 1 - \alpha \of{E} } I_{\text{obs}} \of{E}   }{ \alpha \of{E} } \leq \frac{I_{\text{obs}} \of{E}}{1 -  \alpha \of{E}}.
\end{align}

\begin{figure}[!h]
    \centering
    \includegraphics[width=8.5cm]{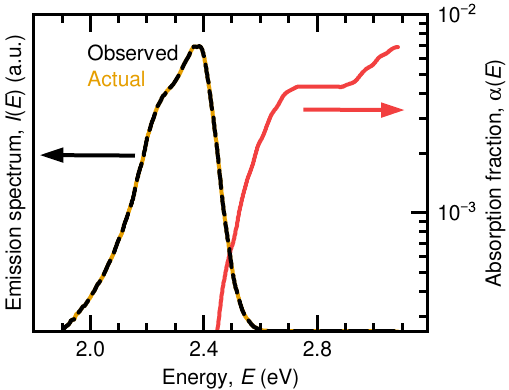}
    \caption{Red line: absorption fraction of a 50 nm thin film of 7 wt\% {\emit} embedded in \host, based on absorption experiments of {\emit} and {\host} in solution \cite{jasparsSpectroElectrochemicalDetermination2021}. Black dashed line: measured emission spectrum of {\emit} in the thin film \cite{jasparsSpectroElectrochemicalDetermination2021}. Orange line: emission spectrum corrected for self-absorption.}
    \label{fig:AutoAbsorption}
\end{figure}

As the emitter {\emit} is embedded in a matrix, the total molar absorption coefficient is a weighted sum of the individual molar absorption coefficients, $ \epsilon \of{E} = \sum_{i} m_i   \epsilon_{i}\of{E} $, where $m_i$ is the mole fraction of species $i$. In Figure~\ref{fig:AutoAbsorption} the absorption fraction is plotted for a $50$ nm thin film with $7$ wt\% ($9$ mol\%) {\emit} (red line). The molar absorption coefficients of {\emit} and {\host} were obtained from Ref.~\cite{jasparsSpectroElectrochemicalDetermination2021}. The resulting emission spectrum is presented in Figure~\ref{fig:AutoAbsorption} (orange line).
We find that the change in the spectrum is negligible. Since only a very small fraction of the photons is reabsorbed (less than 1\%), the change in the spectrum will also be negligible without the simplifying assumptions that we made. 

\end{document}